\documentclass{article}
\usepackage{spconf,amsmath,graphicx}

\usepackage{cite}

\usepackage{amssymb}
\usepackage{xcolor}
\usepackage{hhline}

\usepackage{hyperref}
\usepackage{url}

\usepackage{multirow} 
\usepackage{booktabs}

\title{ON TRAINING TARGETS AND OBJECTIVE FUNCTIONS FOR DEEP-LEARNING-BASED AUDIO-VISUAL SPEECH ENHANCEMENT}
\name{Daniel Michelsanti$^1$, Zheng-Hua Tan$^1$, Sigurdur Sigurdsson$^2$, Jesper Jensen$^{1,2}$}
\address{$^1$ Aalborg University, Department of Electronic Systems, Denmark\\
	$^2$ Oticon A/S, Denmark\\
	\{danmi,zt,jje\}@es.aau.dk $\;$\{ssig,jesj\}@oticon.com}
\begin{document}
\ninept
\maketitle
\begin{abstract}
Audio-visual speech enhancement (AV-SE) is the task of improving speech quality and intelligibility in a noisy environment using audio and visual information from a talker. Recently, deep learning techniques have been adopted to solve the AV-SE task in a supervised manner. In this context, the choice of the target, i.e. the quantity to be estimated, and the objective function, which quantifies the quality of this estimate, to be used for training is critical for the performance. This work is the first that presents an experimental study of a range of different targets and objective functions used to train a deep-learning-based AV-SE system. The results show that the approaches that directly estimate a mask perform the best overall in terms of estimated speech quality and intelligibility, although the model that directly estimates the log magnitude spectrum performs as good in terms of estimated speech quality.

\end{abstract}
\section{Introduction}
\label{sec:intro}

\begin{table*}[t]
\centering
\caption{Objective functions of the approaches used in this study organised according to our taxonomy. Here, $a = \frac{1}{TF}$ and $b = \frac{1}{TQ}$.}
\label{tab:SE-taxonomy}
\resizebox{0.98\textwidth}{!}{%

\begin{tabular}{l | c | c | c }
\toprule

& {Direct Mapping (DM)} & {Indirect Mapping (IM)} & {Mask Approximation (MA)}\\

\midrule

\multirow{ 1}{*}{\rotatebox[origin=c]{90}{\parbox[][0.3cm][c]{0.57cm}{STSA}}} 
& \parbox[][1.2cm][c]{5.47cm}{\begin{align}\tag{1} J = a \sum_{k,l} \big( A_{k,l}- \widehat{A}_{k,l}\big)^2 \label{eqn:STSA-DM}\stepcounter{equation} \end{align}}
& \parbox[][1.2cm][c]{6.12cm}{\begin{align}\tag{6} J = a \sum_{k,l} \big( A_{k,l}- \widehat{M}_{k,l} R_{k,l}\big)^2 \label{eqn:STSA-IM}\stepcounter{equation} \end{align} }
& \parbox[][1.2cm][c]{4.52cm}{\begin{align}\tag{11} J = a \sum_{k,l} \big( M^{\text{IAM}}_{k,l} - \widehat{M}_{k,l}\big)^2 \label{eqn:STSA-MA}\stepcounter{equation} \end{align} }        \\ 

\multirow{ 1}{*}{\rotatebox[origin=c]{90}{\parbox[][0.3cm][c]{0.48cm}{LSA}}} 
{} 
& \parbox[][1.2cm][c]{5.47cm}{\begin{align}\tag{2} J = a\sum_{k,l} \big( \log( A_{k,l} )- \log( \widehat{A}_{k,l} ) \big)^2 \label{eqn:LSA-DM}\stepcounter{equation} \end{align} }         
& \parbox[][1.2cm][c]{6.12cm}{\begin{align}\tag{7} J = a \sum_{k,l} \big( \log( A_{k,l} )- \log( \widehat{M}_{k,l} R_{k,l} ) \big)^2 \label{eqn:LSA-IM}\stepcounter{equation} \end{align} } 
&   -       \\ 

\multirow{ 1}{*}{\rotatebox[origin=c]{90}{\parbox[][0.3cm][c]{0.52cm}{MSA}}} 
{} 
& \parbox[][1.2cm][c]{5.47cm}{\begin{align}\tag{3} J = b \sum_{q,l} \big( \overline{A}_{q,l} - \widehat{\overline{A}}_{q,l} \big)^2 \label{eqn:MSA-DM}\stepcounter{equation} \end{align} }         
& \parbox[][1.2cm][c]{6.12cm}{\begin{align}\tag{8} J = b \sum_{q,l} \big( \overline{A}_{q,l} - \widehat{\overline{M}}_{q,l} \overline{R}_{q,l} \big)^2 \label{eqn:MSA-IM}\stepcounter{equation} \end{align} } 
&    -      \\ 

\multirow{ 1}{*}{\rotatebox[origin=c]{90}{\parbox[][0.3cm][c]{0.6cm}{LMSA}}} 
{} 
& \parbox[][1.2cm][c]{5.47cm}{\begin{align}\tag{4} J = b \sum_{q,l} \big( \log( \overline{A}_{q,l} ) - \log( \widehat{\overline{A}}_{q,l} ) \big)^2 \label{eqn:LMSA-DM}\stepcounter{equation} \end{align} }        
& \parbox[][1.2cm][c]{6.12cm}{\begin{align}\tag{9} J = b \sum_{q,l} \big( \log( \overline{A}_{q,l} ) - \log( \widehat{\overline{M}}_{q,l} \overline{R}_{q,l} ) \big)^2 \label{eqn:LMSA-IM}\stepcounter{equation} \end{align} } 
&  -       \\ 

\multirow{ 1}{*}{\rotatebox[origin=c]{90}{\parbox[][0.3cm][c]{0.54cm}{PSSA}}} 
{} 
& \parbox[][1.2cm][c]{5.47cm}{\begin{align}\tag{5} J = a \sum_{k,l} \big( A_{k,l}\cos(\theta_{k,l}) - \widehat{A}_{k,l} \big)^2 \label{eqn:PSSA-DM}\stepcounter{equation} \end{align} }        
& \parbox[][1.2cm][c]{6.12cm}{\begin{align}\tag{10} J = a \sum_{k,l} \big( A_{k,l}\cos(\theta_{k,l}) - \widehat{M}_{k,l} R_{k,l} \big)^2 \label{eqn:PSSA-IM}\stepcounter{equation} \end{align} } 
& \parbox[][1.2cm][c]{4.52cm}{\begin{align}\tag{12} J = a \sum_{k,l} \big( M^{\text{PSM}}_{k,l} - \widehat{M}_{k,l}\big)^2 \label{eqn:PSSA-MA}\stepcounter{equation} \end{align} }        \\ 

\bottomrule

\end{tabular}
}
\end{table*}

Human-human and human-machine interaction that involves speech as a communication form can be affected by acoustical background noise, which may have a strong impact on speech quality and speech intelligibility. The improvement of one or both of these two speech aspects is known as speech enhancement (SE). Traditionally, this problem has been tackled by adopting audio-only SE (AO-SE) techniques \cite{loizou2013speech, wang2018supervised}. However, speech communication is generally not a unimodal process: visual cues play an important role in speech perception, since they can improve or even alter how phonemes are perceived \cite{mcgurk1976hearing}. This suggests that integrating auditory and visual information can lead to a general improvement in the performance of SE systems. This intuition has lead to the proposal of several audio-visual SE (AV-SE) techniques, e.g. \cite{almajai2011visually}, including deep-learning-based approaches \cite{gabbay2017visual, ephrat2018looking, afouras2018conversation}.

When supervised learning-based methods are used either for AV-SE or for AO-SE, the choice of the target and the objective function used to train the model has a crucial impact on the performance of the system. In this paper, \textit{training target} denotes the desired output of a supervised learning algorithm, e.g. a neural network (NN), while \textit{objective function}, or \textit{cost function}, is the function that quantifies how close the algorithm output is to the target. The effect that targets and objective functions have on AO-SE has been investigated in several works \cite{wang2014training, weninger2014discriminatively, erdogan2015phase}. The estimation of a \textit{mask}, which is used to reconstruct the target speech signal by an element-wise multiplication with a time-frequency (TF) representation of the noisy signal, is usually preferred to a direct estimation of a TF representation of the clean speech signal \cite{erdogan2017deep}. The reason is that a mask is easier to estimate \cite{erdogan2017deep}, because  it is generally smoother than a spectrogram, its values have a narrow dynamic range \cite{wang2014training}, and also because a filtering approach is considered less challenging than the synthesis of a clean spectrogram \cite{afouras2018conversation}. Since no studies on this matter have been performed in the AV domain, design choices of AV frameworks \cite{afouras2018conversation, ephrat2018looking} and their performance \cite{ephrat2018looking} are often motivated by the findings in the AO related works. However, these findings may be inappropriate in the AV domain because, especially at very low signal to noise ratios (SNRs), the estimation of the target is mostly driven by the visual component of the speech. Hence, there is a need for a comprehensive study of the role of training targets and cost functions in AV-SE. 

The contribution of this paper is two-fold. First, we propose a new taxonomy that unifies the different terminologies used in the literature, from classical statistical model-based schemes to more recent deep-learning-based ones. Furthermore, we present a comparison of several targets and objective functions to understand if a particular training target that performs universally good (across various acoustic situations) exists, and if training targets that are good in the AO domain remain good in the AV domain.




\section{Training targets and objective functions}
\label{sec:training targets and objective functions}

Recent works on AO-SE \cite{wang2014training, erdogan2015phase, sun2017multiple} make use of different terminologies for the same approaches. Sometimes, this lack of uniformity can be confusing. In this section, we review cost functions and training targets from the AO domain and introduce a new taxonomy for SE, unifying the terminology used for the classical SE optimisation criteria \cite{ephraim1984speech, ephraim1985speech} and for the objective functions adopted in the recent deep-learning-based techniques \cite{wang2014training, erdogan2015phase} (cf. Table \ref{tab:SE-taxonomy}).

The problem of SE is often formulated as the task of estimating the clean speech signal $x(n)$ given the mixture $y(n)=x(n)+d(n)$, where $d(n)$ is an additive noise signal, and $n$ denotes a discrete-time index. We can formulate the signal model also in the TF domain, as: $Y(k,l)=X(k,l)+D(k,l)$, where $k$ indicates the frequency bin index, $l$ denotes the time frame index, and $Y(k,l)$, $X(k,l)$, and $D(k,l)$ are the short-time Fourier transform (STFT) coefficients of the mixture, the clean signal, and the noise, respectively. Since the STFTs' phases do not have a clear structure, their estimation is hard to perform with a NN \cite{williamson2016complex}. Hence, generally, only the magnitude of the clean STFT is estimated, and the clean signal is reconstructed using the phase of $Y(k,l)$ \cite{wang2014training, erdogan2015phase}.

\subsection{Direct mapping}

Let $A_{k,l}=\mathopen| X(k,l) \mathclose|$ and $R_{k,l}=\mathopen| Y(k,l) \mathclose|$ denote the magnitude of the clean and the noisy STFT coefficients, respectively.  A straightforward way to estimate the short-time spectral amplitude (STSA) of the clean signal is a direct mapping (DM) approach \cite{sun2017multiple}, in which a NN is trained to output an estimate $\widehat{A}_{k,l}$ that minimises a cost function, e.g. Eq. (\ref{eqn:STSA-DM}) {\cite{ephraim1984speech, park2016fully}}, with $k = 1, \ldots , F$ and $l = 1, \ldots , T$, where $F$ is the number of frequency bins of the spectrum estimated by the NN, and $T$ is the number of time frames.
%
%

Since a logarithmic law reflects better the human loudness perception \cite{zwicker2013psychoacoustics}, a cost function that operates in the log spectral amplitude (LSA) domain may be formulated as in Eq. (\ref{eqn:LSA-DM}) {\cite{ephraim1985speech, xu2014experimental}}.
%

To incorporate the fact that the human auditory system is more discriminative at low than at high frequencies \cite{stevens1937scale}, a Mel-scaled spectrum may be defined as $\overline{A}_l = B A_l$, where $A_l$ denotes an $F$-dimensional vector of STFT coefficient magnitudes for time frame $l$, and $B \in \mathbb{R}^{Q \times F}$ is a matrix, implementing a Mel-spaced filter bank, with $Q$ being the number of the Mel-frequency bins. We denote the $q$-th coefficient of the Mel-scaled spectrum at frame $l$ of the clean signal as $\overline{A}_{q,l}$, and its estimate as $\widehat{\overline{A}}_{q,l}$. Then, a cost function in the Mel-scaled spectral amplitude (MSA) domain can be defined as in Eq. (\ref{eqn:MSA-DM})  \cite{lu2013speech}.
%
%

We can combine the considerations leading to Eqs.\ (\ref{eqn:LSA-DM}) and (\ref{eqn:MSA-DM}) to find an estimate that minimises a cost function in the log Mel-scaled spectral amplitude (LMSA) domain, as in Eq. (\ref{eqn:LMSA-DM}) \cite{deng2004enhancement, gabbay2017visual}.
%

Considering only the STSA of the clean signal for the estimation can lead to an inaccurate complex STFT estimation, since the phase of $X(k,l)$ is, generally, different from the phase of $Y(k,l)$ \cite{erdogan2017deep}. For this reason, in \cite{erdogan2015phase}, a factor to compensate for the phase mismatch\footnote{In \cite{erdogan2015phase} a phase compensation factor is used to learn a mask, cf. Eq.\ (\ref{eqn:PSSA-IM}).} is proposed. The cost function that makes use of a phase sensitive spectral amplitude (PSSA) is defined in Eq. (\ref{eqn:PSSA-DM}), where $\theta_{k,l}$ denotes the phase difference between the noisy and the clean signals.
%
%

\subsection{Indirect mapping}

An alternative approach is to have a different training target, and perform an indirect mapping (IM) \cite{weninger2014discriminatively, erdogan2015phase, sun2017multiple}, where a NN is trained to estimate a mask, which is easier to estimate  \cite{erdogan2017deep}, using an objective function which is defined based on reconstructed spectral amplitudes. The cost functions analogous to Eqs.\ (\ref{eqn:STSA-DM})--(\ref{eqn:PSSA-DM}) are defined in Eqs.\ (\ref{eqn:STSA-IM})--(\ref{eqn:PSSA-IM}), where $\widehat{M}_{k,l}$  is the estimate of the magnitude mask, $\widehat{\overline{M}}_{q,l}$ is the estimate of the Mel-scaled mask, and $\overline{R}_{q,l}$ is the Mel-spectrum in frequency subband $q$ and frame $l$ of the noisy signal.


\subsection{Mask approximation}

Since in the IM approach a NN learns a mask, one can also define an objective function directly in the mask domain and perform a mask approximation (MA). In the literature, many different masks have been defined, but in this work we only consider the ideal amplitude mask (IAM), $M^{\text{IAM}}_{k,l}~=~\frac{A_{k,l}}{R_{k,l}}$, and the phase sensitive mask (PSM), $M^{\text{PSM}}_{k,l}~=~\frac{A_{k,l}}{R_{k,l}}\cos(\theta_{k,l})$, because they appear to be the best-performing and allow us to directly compare with the respective IM versions, cf. Eqs.~(\ref{eqn:STSA-IM}) and (\ref{eqn:PSSA-IM}). The cost functions are defined in Eqs.~(\ref{eqn:STSA-MA}) and (\ref{eqn:PSSA-MA})  \cite{wang2014training, erdogan2017deep}, respectively.

%


\noindent While Eqs.\ (\ref{eqn:STSA-MA}) and (\ref{eqn:PSSA-MA}) have led to good performance in the AO-SE domain \cite{wang2014training, williamson2016complex}, the cost functions have been proposed on a heuristic basis. To get insights into their operation, we can rewrite Eq.~(\ref{eqn:STSA-MA}) as $J = \frac{1}{TF} \sum_{k,l} \frac{\big( A_{k,l}  - \widehat{M}_{k,l}{R_{k,l}}\big)^2}{{R_{k,l}}^2}$, which differs from Eq.~(\ref{eqn:STSA-IM}) only due to the $\frac{1}{{R_{k,l}}^2}$ factor. Hence, Eq.~(\ref{eqn:STSA-MA}) is nothing more than a spectrally weighted version of Eq.~(\ref{eqn:STSA-IM}) \cite{fingscheidt2008environment}, which reduces the cost of estimation errors at high-energy spectral regions of the noisy signal relative to low-energy spectral regions, and is related to a perceptually motivated cost function proposed in \cite{loizou2005speech}. Similar considerations can be done for Eqs.~(\ref{eqn:PSSA-IM}) and (\ref{eqn:PSSA-MA}), leading to the conclusion that Eq.~(\ref{eqn:PSSA-MA}) is a spectrally weighted version of Eq.~(\ref{eqn:PSSA-IM}). For simplicity, we refer to the approaches that estimate the IAM and the PSM as STSA-MA and PSSA-MA, respectively.

\section{Experiments}
\label{sec:experiments}

\subsection{Audio-visual corpus and noise data}
\label{subsec:audio-visual corpus and noise data}

We conducted experiments on the GRID corpus \cite{cooke2006audio}, consisting of audio and video recordings of 1000 six-word utterances spoken by each of 34 talkers (s1${-}$34). Each video consists of 75 frames recorded at 25 frames per second with a resolution of 720$\times$576 pixels. The audio tracks have a sample frequency of 44.1 kHz. To train our models, we divided the data as follows: 600 utterances of 25 speakers for training; 600 utterances of 2 speakers (s14 and s15) not in the training set for validation; 25 utterances of each of the speakers in the training set for testing the models in a seen speaker setting; 100 utterances of 6 speakers (s1${-}$4, s7, and s11, 3 males and 3 females) not in the training set for testing the models in an unseen speaker setting. The utterances have been randomly chosen among the ones for which the mouth was successfully detected with the approach described in Sec.~\ref{subsec:audio preprocessing}. 

Six kinds of additive noise have been used in the experiments: bus (BUS), cafeteria (CAF), street (STR) pedestrian (PED), babble (BBL), and speech shaped noise (SSN) as in \cite{kolboek2016speech}. For the training and the validation sets, we mixed the first five noise types with the clean speech signals at 9 different SNRs, in uniform steps between $-20$ dB and $20$ dB. We included SSN in the test set, for the evaluation of the generalisation performance to unseen noise, and evaluated the models between $-15$ dB and $15$ dB SNRs (the performance at $-20$ dB and $20$ dB can be found in \cite{supp_mat}, omitted here due to space limitations). The noise signals used to generate the mixtures in the training, the validation, and the test sets are disjoint over the 3 sets.

\subsection{Audio and video preprocessing}
\label{subsec:audio preprocessing}

Each audio signal was downsampled to 16~kHz and peak-normalised to 1. A TF representation was obtained by applying a 640-point STFT to the waveform signal, using a 640-sample Hamming window and a hop size of 160 samples. The magnitude spectrum was then split into 20-frame-long parts, corresponding to 200 ms, the duration of 5 video frames. Due to spectral symmetry, only the 321 frequency bins that cover the positive frequencies were taken into account.


For each video signal, we first determined a bounding box containing the mouth with the Viola-Jones detection algorithm \cite{viola2001rapid}, and, inside that, we extracted feature points as in \cite{shi1994good} and tracked them across all the video frames using the Kanade-Lucas-Tomasi (KLT) algorithm \cite{lucas1981iterative,tomasi1991detection}. Then, we cropped a mouth-centred region of size 128$\times$128 pixels based on the tracked feature points, and we concatenated 5 consecutive grayscale frames, corresponding to 200 ms.

\subsection{Architecture and training procedure}
\label{subsec:architecture}

Inspired by \cite{gabbay2017visual}, we used a NN architecture that operates in the STFT domain. The NN consists of a video encoder, an audio encoder, a feature fusion subnetwork, and an audio decoder.

The video encoder takes as input 5 frames of size 128$\times$128 pixels obtained as described before, and processes them with 6 convolutional layers, each of them followed by: leaky-ReLU activation, batch normalisation, 2$\times$2  strided max-pooling with kernel of size 2$\times$2, and dropout with a probability of 25\%. Also for the audio encoder, 6 convolutional layers are adopted, followed by leaky-ReLU activation and batch normalisation. The details of the convolutional layers used for the two encoders can be found in \cite{supp_mat}. The input of the audio encoder is a 321$\times$20 spectrogram of the noisy speech signal. Both the audio and video inputs were normalised to have zero mean and unit variance based on the statistics of the full training set.

The two feature vectors obtained as output of the video and the audio encoders are concatenated and used as input to 3 fully-connected layers, the first two having 1312 elements, and the last one 3840 elements. A leaky-ReLU is used as activation function for all the layers. The obtained vector is reshaped to the size of the audio encoder output, and fed into the audio decoder, which has 6 transposed convolutional layers that mirror the layers of the audio encoder. To avoid that the information flow is blocked by the network bottleneck, three skip connections \cite{ronneberger2015u} between the layers 1, 3, and 5 of the audio encoder and the corresponding mirrored layers of the decoder are added to the architecture. A ReLU output layer is applied when the target can assume only positive values (i.e. for all the IM and MA approaches except PSSA-IM and PSSA-MA), otherwise, a linear activation function is used. We clipped the target values between 0 and 10 for the IAM \cite{wang2014training}, and between -10 and 10 for the PSM. The NN outputs a 321$\times$20 spectrogram or a mask.

The networks' weights were initialised with the Xavier approach. For training, we used the Adam optimiser with the objectives previously described. The batch size has been set to 64 and the initial learning rate to $4\cdot10^{-4}$. The NN was evaluated on the validation set every 2 epochs: if the validation loss increased, then the learning rate was decreased to 50\% of its current value. An early stopping technique was adopted: if the validation error did not decrease for 10 epochs, the training was stopped and the model that performed the best on the validation set was used for testing.

\subsection{Audio-visual enhancement and waveform reconstruction}
\label{subsec:enhancement}

To perform the enhancement of a noisy speech signal, we first applied the preprocessing described in Sec.\ \ref{subsec:audio preprocessing} and forward propagated the non-overlapping audio and video segments through the NN. The outputs were concatenated to obtain the enhanced spectrogram of the full speech signal. If the output of the NN was a mask, then the enhanced spectrogram was obtained as the point-wise product between the mask and the spectrogram of the mixture. Finally, the inverse STFT was applied to reconstruct the time-domain signal using the noisy phase. 

\subsection{Evaluation and experimental setup}

The performance of the models was evaluated in terms of perceptual evaluation of speech quality (PESQ) \cite{rix2001perceptual}, as implemented in \cite{loizou2013speech}, and extended short-time objective intelligibility (ESTOI) \cite{jensen2016algorithm}. These metrics have proven to be good estimators of speech quality and intelligibility, respectively, for the noise types considered here.

We designed our experiments to evaluate the approaches listed in Table \ref{tab:SE-taxonomy} in a range of different situations: seen and unseen speaker settings; seen and unseen noise types; different SNRs.

To have a fair comparison for the objective functions, we used the same NN architecture, cf. Sec.\ \ref{subsec:architecture}, and the same input, i.e. a 20-frame-long amplitude spectrum sequence, for all the approaches. The output of the NN always has the same size and can be a magnitude spectrum or a mask to be applied to the noisy spectral amplitudes in the linear domain. When the objective function required the computation of the Mel-scaled spectrum, 80 Mel-spaced frequency bins from 0 to 8 kHz are used \cite{gabbay2017visual}.

For the DM approaches, an exponential function, which can be interpreted as a particular activation function, is applied to the NN output to impose a logarithmic compression of the output values. This makes the dynamic range narrower improving convergence behaviour during training \cite{wang2014training}. No logarithmic compression is applied to PSSA-DM, because PSSA can assume negative values.

 \begin{table*}[t]
\centering
\caption{Results in terms of PESQ and ESTOI. The values are averaged across all the six noise types. The \textit{Unproc.}~rows refer to the unprocessed signals, and the \textit{AO} columns show the average scores for models without the video encoder, trained only on the audio signals.}
\label{tab:results}
\resizebox{0.98\textwidth}{!}{%

\begin{tabular}{l || ccccccc | cc || ccccccc | cc}

\toprule

 PESQ & & \multicolumn{7}{c}{Seen Speakers} & & \multicolumn{9}{c}{Unseen Speakers} \\ 
 
 \midrule

SNR (dB) & -15 & -10 & -5 & 0 & 5 & 10 & 15 & Avg. & AO & -15 & -10 & -5 & 0 & 5 & 10 & 15 & Avg. & AO \\ 

\midrule

Unproc.  
& 1.09 & 1.08 & 1.08 & 1.11 & 1.20 & 1.39 & 1.71 & 1.24 & 1.24 
& 1.10 & 1.09 & 1.08 & 1.11 & 1.20 & 1.39 & 1.70 & 1.24 & 1.24
\\

\midrule

STSA-DM
 & \textbf{1.27} & 1.35 & 1.48 & 1.65 & 1.86 & 2.08 & 2.31 & 1.71 & 1.59
 & 1.13 & 1.19 & 1.30 & 1.48 & 1.73 & 1.99 & 2.24 & 1.58 & 1.57
\\ 
LSA-DM   
 & 1.24 & 1.37 & \textbf{1.57} & \textbf{1.84} & \textbf{2.14} & \textbf{2.45} & \textbf{2.74} & \textbf{1.91} &  \textbf{1.74} 
 & \textbf{1.15} & \textbf{1.23} & \textbf{1.37} & \textbf{1.59} & \textbf{1.91} & \textbf{2.25} & \textbf{2.57} & \textbf{1.72} & \textbf{1.70}
\\ 
MSA-DM  
 & \textbf{1.27} & 1.36 & 1.49 & 1.67 & 1.87 & 2.07 & 2.28 & 1.72 & 1.58
 & 1.14 & 1.20 & 1.32 & 1.51 & 1.75 & 1.99 & 2.21  & 1.59 & 1.56
\\ 
LMSA-DM
 & \textbf{1.27} & \textbf{1.39} & 1.56 & 1.78 & 2.01 & 2.18 & 2.31 & 1.79 &  1.62
 & \textbf{1.15} & 1.22 & 1.34 & 1.53 & 1.77 & 1.98 & 2.14 & 1.59 & 1.59
\\ 
PSSA-DM
 & 1.24 & 1.32 & 1.44 & 1.61 & 1.82 & 2.04	 & 2.25 & 1.67 & 1.62
 & 1.13 & 1.18 & 1.28 & 1.45 & 1.70 & 1.94 & 2.17   & 1.55 & 1.58
\\

\midrule

STSA-IM
 & 1.24 & 1.33 & 1.45	& 1.61 & 1.77 & 1.95	 & 2.19 & 1.65 &  1.58
 & 1.13 & 1.18 & 1.28 & 1.44 & 1.65 & 1.87 & 2.11 & 1.52 & 1.56
\\ 
LSA-IM    
& 1.17 & 1.25 & 1.39 & 1.60 & 1.89 & 2.19 & 2.49 & 1.71 & 1.57
& 1.13 & 1.17 & 1.28 & 1.46 & 1.72 & 2.02 & 2.34  & 1.59 & 1.57
\\ 
MSA-IM   
& 1.26 & 1.34 & 1.47 & 1.64 & 1.85 & 2.07 & 2.30 & 1.70 & \textbf{1.65}
& 1.13 & 1.19 & 1.29 & 1.47 & 1.71 & 1.98 & 2.24  & 1.57 & \textbf{1.63}
\\ 
LMSA-IM  
 & 1.21 & 1.32 & 1.48 & \textbf{1.72} & \textbf{1.99} & \textbf{2.26} & \textbf{2.53} & \textbf{1.79} &  1.56
& 1.13 & 1.19 & 1.30 & 1.49 & \textbf{1.76} & \textbf{2.06} & \textbf{2.35} & \textbf{1.61} & 1.55
\\ 
PSSA-IM  
& \textbf{1.29} & \textbf{1.37} & \textbf{1.50} & 1.68 & 1.87 & 2.05	 & 2.22 & 1.71 & \textbf{1.65}
& \textbf{1.16} & \textbf{1.22} & \textbf{1.33} & \textbf{1.51} & 1.74 & 1.96 & 2.15 & 1.58 & 1.62
\\

\midrule

STSA-MA    
& \textbf{1.31} & \textbf{1.42} & \textbf{1.57} & \textbf{1.78} & 2.02 & 2.29 & 2.58 & 1.85 &  1.62  
& 1.15 & 1.21 & 1.32 & 1.52 & 1.81 & 2.15 & 2.48 & 1.66 & 1.62
\\ 

PSSA-MA  
& 1.28 & 1.38 & 1.54 & \textbf{1.78} & \textbf{2.08} & \textbf{2.40} & \textbf{2.71} & \textbf{1.88} & \textbf{1.77} 
& \textbf{1.18} & \textbf{1.25} & \textbf{1.38} & \textbf{1.61} & \textbf{1.95} & \textbf{2.31} & \textbf{2.63} & \textbf{1.76} & \textbf{1.76}
\\

\bottomrule

\end{tabular}%
}

\resizebox{0.98\textwidth}{!}{%

\begin{tabular}{l || ccccccc | cc || ccccccc | cc}

\toprule

 ESTOI & & \multicolumn{7}{c}{Seen Speakers} & & \multicolumn{9}{c}{Unseen Speakers} \\ 
 
 \midrule

SNR (dB) & -15 & -10 & -5 & 0 & 5 & 10 & 15 & Avg. & AO & -15 & -10 & -5 & 0 & 5 & 10 & 15 & Avg. & AO \\ 

\midrule

Unproc.  
&0.08&0.15&0.24&0.35&0.47&0.58&0.67 & 0.36 & 0.36
& 0.08 & 0.14 & 0.23 & 0.34 & 0.46 & 0.57 & 0.66 & 0.35 & 0.35
\\

\midrule

STSA-DM
& 0.35& 0.41& 0.49	& 0.57& 0.64& 0.70& 0.74 & 0.56 & 0.48
& 0.23 & 0.29 & 0.39 & 0.49 & 0.59 & 0.67 & 0.72 & 0.48 & \textbf{0.47}
\\ 
LSA-DM   
& 0.35& 0.41& 0.49	& 0.58& 0.65& \textbf{0.71}& \textbf{0.76} & 0.56 & 0.48
& 0.24 & 0.30 & 0.39 & 0.49 & 0.60 & \textbf{0.68} & \textbf{0.73} & 0.49 & \textbf{0.47}
\\ 
MSA-DM  
& 0.36& 0.42& 0.49& 0.57& 0.64& 0.70& 0.74 & 0.56 & \textbf{0.49}
& 0.24 & \textbf{0.31} & \textbf{0.40} & \textbf{0.51} & \textbf{0.61} & \textbf{0.68} & \textbf{0.73} & \textbf{0.50} & \textbf{0.47}
\\ 
LMSA-DM
& \textbf{0.37}& \textbf{0.44}& \textbf{0.51}& \textbf{0.60}& \textbf{0.66}& \textbf{0.71}& 0.75 & \textbf{0.58} & 0.48
& \textbf{0.25} & \textbf{0.31} & \textbf{0.40} & \textbf{0.51} & \textbf{0.61} & \textbf{0.68} & 0.72 & \textbf{0.50} & 0.46
\\ 
PSSA-DM
& 0.29& 0.36& 0.46& 0.56& 0.64& 0.70& 0.74 & 0.53 & \textbf{0.49}
& 0.19 & 0.27 & 0.37 & 0.49 & 0.60 & \textbf{0.68} & 0.72 & 0.48 & \textbf{0.47}
\\

\midrule

STSA-IM
& 0.33& 0.40& 0.48& 0.56& 0.64& 0.69& 0.74 & 0.55 & 0.49
& 0.23 & 0.29 & 0.39 & \textbf{0.50} & 0.60 & 0.67 & 0.72 & 0.48 & 0.47
\\ 
LSA-IM   
& 0.33& 0.38& 0.46& 0.55& 0.63& 0.70& 0.75  & 0.54 & 0.46
& 0.22 & 0.28 & 0.36 & 0.46 & 0.57 & 0.66 & \textbf{0.73}  & 0.47 & 0.45
\\ 
MSA-IM   
& \textbf{0.36}& \textbf{0.42}& \textbf{0.50}& 0.58& 0.65& 0.70& 0.75 & \textbf{0.57} & \textbf{0.50}
& \textbf{0.25} & \textbf{0.31} & \textbf{0.40} & \textbf{0.50} & 0.60 & \textbf{0.68} & \textbf{0.73} & \textbf{0.50} & \textbf{0.48}
\\ 
LMSA-IM 
& \textbf{0.36}& \textbf{0.42}& \textbf{0.50}& \textbf{0.59}& \textbf{0.66}& \textbf{0.72}& \textbf{0.76}  & \textbf{0.57} & 0.47
& 0.24 & 0.30 & 0.38 & 0.49 & 0.60 & \textbf{0.68} & \textbf{0.73} & 0.49 & 0.46
\\ 
PSSA-IM  
& 0.29& 0.37& 0.46& 0.56& 0.64& 0.70& 0.75 & 0.54 & 0.49
& 0.21 & 0.28 & 0.38 & \textbf{0.50} & \textbf{0.61} & \textbf{0.68} & \textbf{0.73} & 0.48 & 0.47
\\

\midrule

STSA-MA   
& \textbf{0.39}& \textbf{0.45}& \textbf{0.52}& \textbf{0.60}& \textbf{0.67}& \textbf{0.72}& \textbf{0.77}  & \textbf{0.59} & 0.49
& \textbf{0.26} & \textbf{0.32} & \textbf{0.41} & 0.51 & 0.62 & \textbf{0.70} & \textbf{0.75} & \textbf{0.51} &  0.48
\\ 

PSSA-MA  
& 0.29& 0.36& 0.46& 0.57& 0.66& \textbf{0.72}& \textbf{0.77} & 0.55 & \textbf{0.50}
& 0.22 & 0.29 & 0.40 & \textbf{0.52} & \textbf{0.63} & \textbf{0.70} & \textbf{0.75} & 0.50 & \textbf{0.49}
\\

\bottomrule

\end{tabular}%
}

\end{table*}

\section{Results and discussion}
\label{sec:results}

Table \ref{tab:results} shows the results of the experiments. For the seen speaker case (left half of the table), all SE methods clearly improve the noisy signals in terms of both estimated quality and intelligibility. Regarding PESQ, LSA-DM achieves the best results overall, closely followed by the MA approaches. Among the IM techniques, the ones that operate in the log domain are the best at high SNRs, but at low SNRs the phase-aware target appears to be beneficial. There is no big difference in terms of ESTOI among the various methods, however at very low SNRs, the phase sensitive approaches do not perform as well as the other methods. This is surprising, since it was not observed in the AO setting \cite{erdogan2015phase, supp_mat}, and should be investigated further. 
Even though the approaches that operate in the Mel domain seem to have no advantages in terms of PESQ, they allow to achieve slightly higher ESTOI for both DM and IM.

For the unseen speaker case, the behaviour is similar, with small differences among the methods in terms of ESTOI. Regarding PESQ, LSA-DM is the approach showing the largest improvements among the DM ones, and it is slightly worse than PSSA-MA.

A comparison between the seen and the unseen speakers conditions makes it clear that, at very low SNRs, knowledge of the speaker is an advantage: for example, ESTOI values at $-15$ dB SNR for the seen speakers are higher than the ones for the unseen speakers at $-10$ dB. This can be explained by the fact that the speech characteristics of an unseen speaker are harder to reconstruct by the NN, because some information of the voice attributes, e.g. pitch and timbre, cannot be easily derived from the mouth movements only. 

From the results of the AO models, we observe that, generally, visual information helps in improving systems performance. The widest gap between the AV-SE systems and the respective AO-SE ones is reported for the seen speakers case. However, for unseen speakers, we see no significant improvements in terms of estimated speech quality, but for estimated speech intelligibility, the AV models are, on average, slightly better than the respective AO models. The performance difference between AO and AV models is mostly notable at low SNRs, with a gain of about 5 dB (cf. \cite{supp_mat}).



The results for the unseen noise type (SSN) in isolation have not been reported due to space limitations, but can be found in \cite{supp_mat}. All the systems show reasonable generalisation performance to this noise type with an improvement over the noisy signals similar to the one observed for the seen BBL noise type in terms of ESTOI.

Overall, the three best approaches among the ones investigated are LSA-DM, STSA-MA, and PSSA-MA.

\section{Conclusion}
\label{sec:conclusion}

In this study, we proposed a new taxonomy to have a uniform terminology that links classical speech enhancement methods with more recent techniques, and investigated several training targets and objective functions for audio-visual speech enhancement. We used a deep-learning-based framework to directly and indirectly learn the short time spectral amplitude of the target speech in different domains. The mask approximation approaches and the direct estimation of the log magnitude spectrum are the methods that perform the best. In contrast to the results for audio-only speech enhancement, the use of a phase-aware mask is not as effective in improving estimated intelligibility especially at low SNRs.

\bibliographystyle{IEEEbib}
\bibliography{icassp1}

\begin{thebibliography}{10}

\bibitem{loizou2013speech}
P.~C. Loizou,
\newblock {\em Speech enhancement: theory and practice},
\newblock CRC press, 2013.

\bibitem{wang2018supervised}
D.~L. Wang and J.~Chen,
\newblock ``Supervised speech separation based on deep learning: An overview,''
\newblock {\em IEEE/ACM Transactions on Audio, Speech, and Language
  Processing}, 2018.

\bibitem{mcgurk1976hearing}
H.~McGurk and J.~MacDonald,
\newblock ``Hearing lips and seeing voices,''
\newblock {\em Nature}, vol. 264, no. 5588, pp. 746--748, 1976.

\bibitem{almajai2011visually}
I.~Almajai and B.~Milner,
\newblock ``Visually derived {W}iener filters for speech enhancement,''
\newblock {\em IEEE Transactions on Audio, Speech, and Language Processing},
  vol. 19, no. 6, pp. 1642--1651, 2011.

\bibitem{gabbay2017visual}
A.~Gabbay, A.~Shamir, and S.~Peleg,
\newblock ``Visual speech enhancement,''
\newblock in {\em Proc. of Interspeech}, 2018.

\bibitem{ephrat2018looking}
A.~Ephrat, I.~Mosseri, O.~Lang, T.~Dekel, K.~Wilson, A.~Hassidim, W.~T.
  Freeman, and M.~Rubinstein,
\newblock ``Looking to listen at the cocktail party: A speaker-independent
  audio-visual model for speech separation,''
\newblock {\em ACM Transactions on Graphics}, vol. 37, no. 4, pp.
  112:1--112:11, 2018.

\bibitem{afouras2018conversation}
T.~Afouras, J.~S. Chung, and A.~Zisserman,
\newblock ``The conversation: Deep audio-visual speech enhancement,''
\newblock {\em Proc. of Interspeech}, 2018.

\bibitem{wang2014training}
Y.~Wang, A.~Narayanan, and D.~L. Wang,
\newblock ``On training targets for supervised speech separation,''
\newblock {\em IEEE/ACM Transactions on Audio, Speech and Language Processing},
  vol. 22, no. 12, pp. 1849--1858, 2014.

\bibitem{weninger2014discriminatively}
F.~Weninger, J.~R. Hershey, J.~Le~Roux, and B.~Schuller,
\newblock ``Discriminatively trained recurrent neural networks for
  single-channel speech separation,''
\newblock in {\em Proc. of GlobalSIP}, 2014.

\bibitem{erdogan2015phase}
H.~Erdogan, J.~R. Hershey, S.~Watanabe, and J.~Le~Roux,
\newblock ``Phase-sensitive and recognition-boosted speech separation using
  deep recurrent neural networks,''
\newblock in {\em Proc. of ICASSP}, 2015.

\bibitem{erdogan2017deep}
H.~Erdogan, J.~R. Hershey, S.~Watanabe, and J.~Le~Roux,
\newblock ``Deep recurrent networks for separation and recognition of
  single-channel speech in nonstationary background audio,''
\newblock in {\em New Era for Robust Speech Recognition}, pp. 165--186.
  Springer, 2017.

\bibitem{sun2017multiple}
L.~Sun, J.~Du, L.-R. Dai, and C.-H. Lee,
\newblock ``Multiple-target deep learning for {LSTM-RNN} based speech
  enhancement,''
\newblock in {\em Proc. of HSCMA}, 2017.

\bibitem{ephraim1984speech}
Y.~Ephraim and D.~Malah,
\newblock ``Speech enhancement using a minimum-mean square error short-time
  spectral amplitude estimator,''
\newblock {\em IEEE Transactions on Acoustics, Speech, and Signal Processing},
  vol. 32, no. 6, pp. 1109--1121, 1984.

\bibitem{ephraim1985speech}
Y.~Ephraim and D.~Malah,
\newblock ``Speech enhancement using a minimum mean-square error log-spectral
  amplitude estimator,''
\newblock {\em IEEE Transactions on Acoustics, Speech, and Signal Processing},
  vol. 33, no. 2, pp. 443--445, 1985.

\bibitem{williamson2016complex}
D.~S. Williamson, Y.~Wang, and D.~L. Wang,
\newblock ``Complex ratio masking for monaural speech separation,''
\newblock {\em IEEE/ACM Transactions on Audio, Speech and Language Processing},
  vol. 24, no. 3, pp. 483--492, 2016.

\bibitem{park2016fully}
S.~R. Park and J.~Lee,
\newblock ``A fully convolutional neural network for speech enhancement,''
\newblock in {\em Proc. of Interspeech}, 2017.

\bibitem{zwicker2013psychoacoustics}
E.~Zwicker and H.~Fastl,
\newblock {\em Psychoacoustics: Facts and models}, vol.~22,
\newblock Springer Science \& Business Media, 2013.

\bibitem{xu2014experimental}
Y.~Xu, J.~Du, L.-R. Dai, and C.-H. Lee,
\newblock ``An experimental study on speech enhancement based on deep neural
  networks,''
\newblock {\em IEEE Signal processing letters}, vol. 21, no. 1, pp. 65--68,
  2014.

\bibitem{stevens1937scale}
S.~S. Stevens, J.~Volkmann, and E.~B. Newman,
\newblock ``A scale for the measurement of the psychological magnitude pitch,''
\newblock {\em The Journal of the Acoustical Society of America}, vol. 8, no.
  3, pp. 185--190, 1937.

\bibitem{lu2013speech}
X.~Lu, Y.~Tsao, S.i Matsuda, and C.~Hori,
\newblock ``Speech enhancement based on deep denoising autoencoder,''
\newblock in {\em Proc. of Interspeech}, 2013.

\bibitem{deng2004enhancement}
L.~Deng, J.~Droppo, and A.~Acero,
\newblock ``Enhancement of log {Mel} power spectra of speech using a
  phase-sensitive model of the acoustic environment and sequential estimation
  of the corrupting noise,''
\newblock {\em IEEE Transactions on Speech and Audio Processing}, vol. 12, no.
  2, pp. 133--143, 2004.

\bibitem{fingscheidt2008environment}
T.~Fingscheidt, S.~Suhadi, and S.~Stan,
\newblock ``Environment-optimized speech enhancement,''
\newblock {\em IEEE Transactions on Audio, Speech, and Language Processing},
  vol. 16, no. 4, pp. 825--834, 2008.

\bibitem{loizou2005speech}
P.~C. Loizou,
\newblock ``Speech enhancement based on perceptually motivated {Bayesian}
  estimators of the magnitude spectrum,''
\newblock {\em IEEE Transactions on Speech and Audio Processing}, vol. 13, no.
  5, pp. 857--869, 2005.

\bibitem{cooke2006audio}
M.~Cooke, J.~Barker, S.~Cunningham, and X.~Shao,
\newblock ``An audio-visual corpus for speech perception and automatic speech
  recognition,''
\newblock {\em The Journal of the Acoustical Society of America}, vol. 120, no.
  5, pp. 2421--2424, 2006.

\bibitem{kolboek2016speech}
M.~Kolb{\ae}k, Z.-H. Tan, and J.~Jensen,
\newblock ``Speech enhancement using long short-term memory based recurrent
  neural networks for noise robust speaker verification,''
\newblock in {\em Proc. of SLT}, 2016.

\bibitem{supp_mat}
D.~Michelsanti, Z.-H. Tan, S.~Sigurdsson, and J.~Jensen,
\newblock ``On training targets and objective functions for deep-learning-based
  audio-visual speech enhancement - supplementary material,''
  \url{http://kom.aau.dk/~zt/online/icassp2019_sup_mat.pdf}, 2019.

\bibitem{viola2001rapid}
P.~Viola and M.~Jones,
\newblock ``Rapid object detection using a boosted cascade of simple
  features,''
\newblock in {\em Proc. of CVPR}, 2001.

\bibitem{shi1994good}
J.~Shi and C.~Tomasi,
\newblock ``Good features to track,''
\newblock in {\em Proc. of CVPR}, 1994.

\bibitem{lucas1981iterative}
B.~D. Lucas and T.~Kanade,
\newblock ``An iterative image registration technique with an application to
  stereo vision,''
\newblock in {\em Proc. of IJCAI}, 1981.

\bibitem{tomasi1991detection}
C.~Tomasi and T.~Kanade,
\newblock ``Detection and tracking of point features,''
\newblock {\em Technical Report CMU-CS-91-132}, 1991.

\bibitem{ronneberger2015u}
O.~Ronneberger, P.~Fischer, and T.~Brox,
\newblock ``U-net: Convolutional networks for biomedical image segmentation,''
\newblock in {\em Proc. of MICCAI}, 2015.

\bibitem{rix2001perceptual}
A.~W. Rix, J.~G. Beerends, M.~P. Hollier, and A.~P. Hekstra,
\newblock ``Perceptual evaluation of speech quality ({PESQ}) - a new method for
  speech quality assessment of telephone networks and codecs,''
\newblock in {\em Proc. of ICASSP}, 2001.

\bibitem{jensen2016algorithm}
J.~Jensen and C.~H. Taal,
\newblock ``An algorithm for predicting the intelligibility of speech masked by
  modulated noise maskers,''
\newblock {\em IEEE/ACM Transactions on Audio, Speech, and Language
  Processing}, vol. 24, no. 11, pp. 2009--2022, 2016.

\end{thebibliography}

\end{document}